# Predictive Estimation of the Optimal Signal Strength from Unmanned Aerial Vehicle over Internet of Things Using ANN

S. H. Alsamhi[1], Ou Ma[2], M. S. Ansari[3]

[1]School of Aerospace, Tsinghua University, Beijing, China, and IBB University, Ibb, Yemen
[2]College of Engineering and Applied Science, University of Cincinnati, Ohio, USA
[3]Department of Electronics Engineering, AMU, Aligarh, India
salsamhi@tsinghua.edu.cn, ou.ma@uc.edu, samar.ansari@zhcet.ac.in.

**Abstract:** - This paper proposes an intelligent technique for maximizing the network connectivity and provisioning desired quality of service (QoS) of integration of internet of things (IoT) and unmanned aerial vehicle (UAV). Prediction of the signal strength and fading channel conditions enable adaptive data transmission which turn enhances the QoS for the end users/ devices with reducing the power consumption for data transmissions. UAV is data gathering robot from the difficult or impossible area for humans to reach. Hence, Atmospheric dynamics and environment influence the signal strength during traveling in space among UAV, IoT devices, and humankind. Therefore, Signal moving from the smart UAV is sensitive to the effects of attenuation, reflection, diffraction, scattering, and shadowing. Also, we analyze the ability ANN to predictively estimate the signal strength and channel propagation from the drone and physical medium parameters. Moreover, the results show that the distortion of the signal can be reduced and enhanced significantly.

**Keywords:** Signal Strength prediction, ANN, UAV, QoS, Devices connectivity

## I. Introduction

The integration of Artificial Intelligence (AI), communication and robotics may lead mainly to change the world and our living way. Recently, robotic communication became increasingly common for people used in various situations, i.e., home, hospitals, hotels，and industry. To make robots exchange the information with each other as well as with human; the appreciated techniques for communication link should be established. Establishing the communication link among robots and human depends on the environmental impacts. To enhance the communication link, signal strength, channel distribution, elevation angle, and radiation power should be taken into consideration. During transmission signal, the signal strength suffers from dynamics of the atmosphere and the environment impacts such as scattering, diffraction, reflection, and shadowing of electromagnetic waves, which can powerfully distort the signals.

Furthermore, the reliability and the quality of signals from the drones to the ground are affected by the large and small fading. Therefore, the performance of communication link among the drones and other robots on the ground get more degradation and attenuation. Many types of research have been done for developing the communication links of the satellite and terrestrial [1, 2]. However, the unmanned aerial vehicle (UAV) communication links still new, and it is different from the satellite and terrestrial links. The significant difference is the motion UAV (transmitter) in a homogeneous environment where receiver agents located on the ground with different environments. Recently, drones became attractive research and industrial area, because of the flexibility and the possibility of using them in many applications such as security, controlling, monitoring, and exploring the areas difficult to reach.

The most impotrant feature of the UAV are a line of sight (LoS). This feature leads to enhance the signal rate, deliver communication services to large coverage area and deploy to improve the quality of services (QoS), efficiently [3, 4]. Drones can be used for delivering services to any temporary events such as sports, deliver parcel services in a high traffic area, stadiums, traffic monitoring and so on. To perform such tasks, the signal strength should be taken into consideration. Hence, it is our responsibility to find the optimal and accurate intelligence technique for predicting the signal strength among the drone, Internet of Things (IoT) devices, and agents on the ground for designing the receiver and transmitter, accordingly. Therefore, we investigate the impact of a limited number of signal strength measurements on the accuracy of coverage prediction and estimation of propagation parameters.

The rest of this paper is organized as follows. The related work was discussed in Section II, and the UAVs described in Section III. The machinist of signal propagation and the proposed technique are presented in Section IV and V, respectively. Section VI presents the results, and the paper is concluded in Section VII.

## II. Related work

Collecting data from the IoT devices is the most attractive application of drones [5, 6]. The UAV can communicate with IoT devices which are unable to communicate over a large area and send information to the intended receiver as shown in Fig.1. The development of communication technologies leads to change the way we communicate, which a vast number of things are connected to the Internet. Internet of people refers to the Internet connects all of the peoples. However, the connections among things refer to the internet of things (IoT). The concept of IoT refers to things, which are not intelligent and do not include Artificial Intelligence (AI). However, Internet of Robotic Things (IoRT) is an intelligent concept which gives associated things, the ability of negotiation, reasoning, and delegation.

Integrated of IoT devices and robots are growing upon ecosystem, whereas IoT devices, robots, and human communicate on essential cooperative. The target applications and technological implications of IoT- aided robotics were discussed [7]. Furthermore, Dutta et al. [8] addressed the network security enhancement to IoT- aided robotics in the complex environment. Dauphin et al. [9] reviewed the convergence in term of network protocols and embedded software for IoT robotic. The interaction of robotic and IoT devices has been investigated in [10]. Therefore, AI, robots, and IoT will provide the next generation of IoT applications [11].

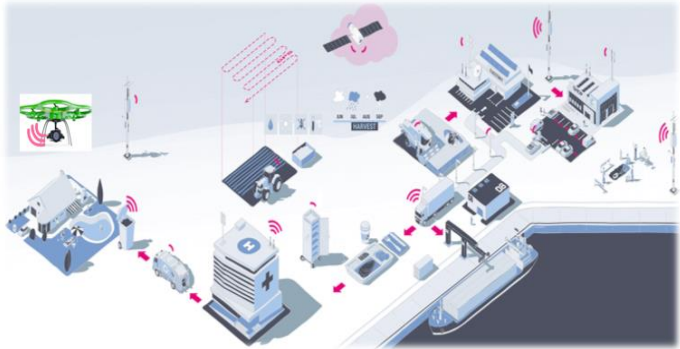

Fig.1 UAV for collecting data from IoT devices

Moreover, Razafimandimby et al. [12] implemented an intelligence technique to preserve global connectivity between IoRT robots by using ANN. ANN played a vital role in the balance between desired network coverage and the QoS of IoRT communication. Internet of Intelligent Things (IoIT) and the robot as a service (RaaS) interfacing within the cloud computing environment are discussed[13]. Architecture, interface, and behavior of RaaS were defined. Furthermore, the dependability design was the focus of the research. The internet of vehicles (IoV) has communication devices, storage, intelligence and learning capabilities to anticipate the customers' intentions [14]. Internet of fly things concept was discussed in [15]. Also, agents provided an efficient mechanism for communication amongst networked heterogeneous devices within a traffic information system [16].

The signal propagates via a dynamic atmosphere play vital role in desired QoS and maintaining the network connectivity. The signal path at the desired frequency effects of various multipath components occurring based on the terrain, small-scale fading, and the impact of Doppler shift due to the movement of the drones, the delay dispersion are characterized. The path loss from drones to the IoT devices on the ground discussed based on elevation angle [17], based on the height of the UAV[18]. Al-Hourani et al. [19] argued the optimal altitude for enabling the signal for maximum coverage area. Statistical propagation for predicting the path loss between UAV and the terrestrial terminal was given in [20]. Bor-Yaliniz et al. [21] envisioned a multi-UAV cellular network which is bringing the delivery of wireless networks to anywhere as per demand at any time.

The prediction of the signal strength is critical to design both transmitter and receiver for any communication system. The artificial intelligence plays crucial strategy for appreciating estimation of signal strength as shown in Fig.2 Regarding indoor environments neuro-fuzzy is proposed to simulate the propagation model for predicting RSS and comparing the performance with empirical models of channels [22]. RSS from the aerial platform is calculated using Hata model [23]. However, Kaiser [24] integrated the combination of received signal strength (RSS) and variance fractal dimensions as an input to ANN for prediction of location features.

Moazzeni. [25] argued the relation between path loss, delay, and another atmosphere environment such as pressure, humidity, and temperature. ANN was used for channel propagation and verified the accuracy and validation. Furthermore, the flexibility and processing speed of ANN was the reason for adapting as an alternative technique for channel propagation and signal strength in various environments [26]. ANN was used to predict channel propagation of multi-user transmission under Rayleigh fading for maximizing the efficiencies [27]. In addition, learned estimators were improved the short-time under fading channel conditions and propagation effects [28]. Also, Bernadó et al. [29] focused and investigated the channel characteristics and variation of K-factor of the non-stationary vehicle to a vehicle (V2V) in time and frequency. Furthermore, Tomasevic et al. [30] observed that the performance of proposed ANN was better agreement than existing simulation schemes.

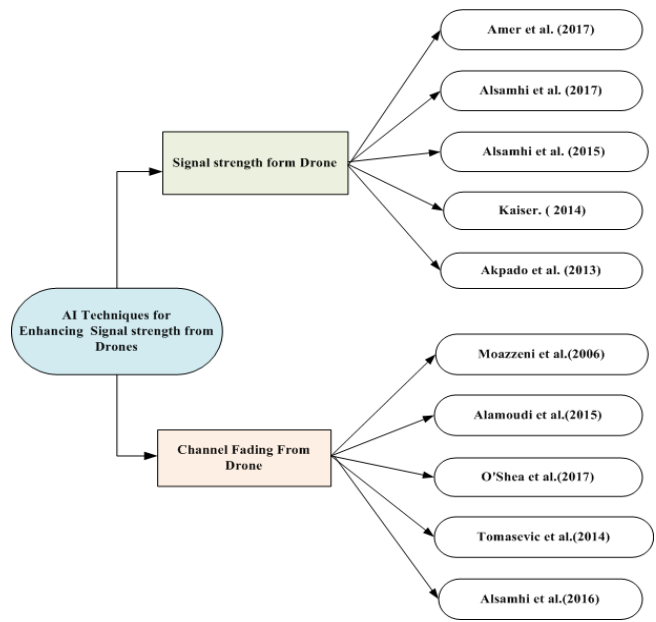

Fig.2 A broad AI Techniques of predicting the signal strength from drone

## III. Integrated smart drones and IoT

Recently, drones play the vital role in developing the next generation of wireless network and IoT. Therefore, it attracts the attention of people to quickly developing UAV applications which serve society, industry, government, and environments. Furthermore, drones have a branch of sensors that are getting real-time information from everywhere they fly. They are already beginning to efficiently replace that connected sensors at stationary with one device which has most essential features such as easy to deploy, flexible payloads, reprogrammable, measure anything at any time in anywhere. The UAV can enable the communication services, while the wireless communication networks are damaged during disaster [31]. It can be used as capture information for a particular place and send in real time to center for taking the proper decision accordingly. The optimal placement of drones for public safety communications to enhance the coverage performance was discussed in [32].

Nowadays, drones became an important part, and the future of the IoT for many reasons：can be deployable, carry a flexible payload, reprogrammable in the mission, not static, can measure anything in anywhere. Furthermore, drones offer several advantages over IoT stationary devices such as used for lengthy periods at preferred altitudes, better resolution image quality, low cost, rapid response, capable of flying without any condition, able to be closer to areas of interest. They represent an emerging and useful technique for delivering wireless service to the user or the IoT devices on the ground which investigated for large coverage area and reduces the power consumption [33].

UAV is one of the aerial robotics which has attracted the interest for broad applications significantly. It is not only vehicle flying in the space, but it is a system which has network architecture including space segment, ground segment and between them the communication link as shown in Fig.3. UAV is used to collect data from IoT devices on the ground, process data, and send data to the destination for further action and decision, accordingly. Jiang et al. [34] discussed the static ground users, the optimal trajectory, and heading of UAVs was equipped with multiple antennas for the ground to air uplink scenario. The reliable channel between UAV and ground stuff is highly required.

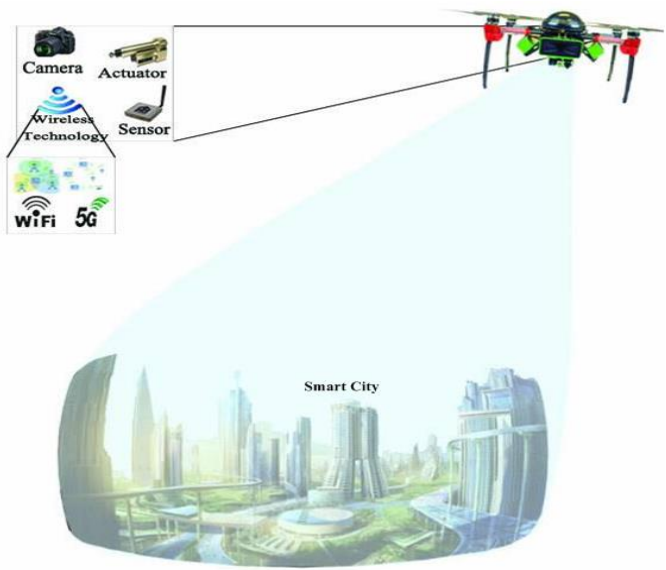

Fig.3 Integrated smart UAV and IoT network structure

A. Sky segment includes vehicle and payload which are suitable for the required systems to perform the tasks efficiently. The payload may contain cameras, sensors, antennas, actuators, a processing unit and wireless components such as ad hoc network, WiFi or LTE.

B. Ground segment: the most critical part is ground control station (GCS) which includes the required equipment for flight planning, mission monitoring, guiding and giving comments. UAV delivers services to the robots/ humans in different environments for different tasks. IoT devices are the part of ground segments which are used to collect data around their environments.

C. The communication link is divided into control and command link. It may be a line of sight (LoS), visual line of sight (VoS), or beyond of sight (BoS).

The applications of UAV are increased in demand of raising the required for extensive research and development of technologies in the fields of smart industry, smart home, smart city, smart environments, smart mobility, smart living, and smart everything. Previously, it was used only for military applications and tasks. However, it can also use in civilian applications such as surveillance, sensing, relief and search in disaster, reconnaissance, observation, aerial surveying, monitoring, remote sensing, and exploration, etc.

## IV. Propagation signal mechanism from UAV to IoT devices

The most important part of the propagation is the study of the signal strength. With the help of signal strength, drones are controlled, keep moving in the particular trajectory, receive the signal from the agents at the ground, follow the drones moving, detect the specific object, and help drones to reach precise position. The traveling signal in space suffers from different phenomena's. These phenomena's effect on the strength of the signal among UAV and agents or IoT devices on the ground; then it will reach the destination very weak. Therefore, these phenomena should be taken into consideration for designing the transmission and receiver.

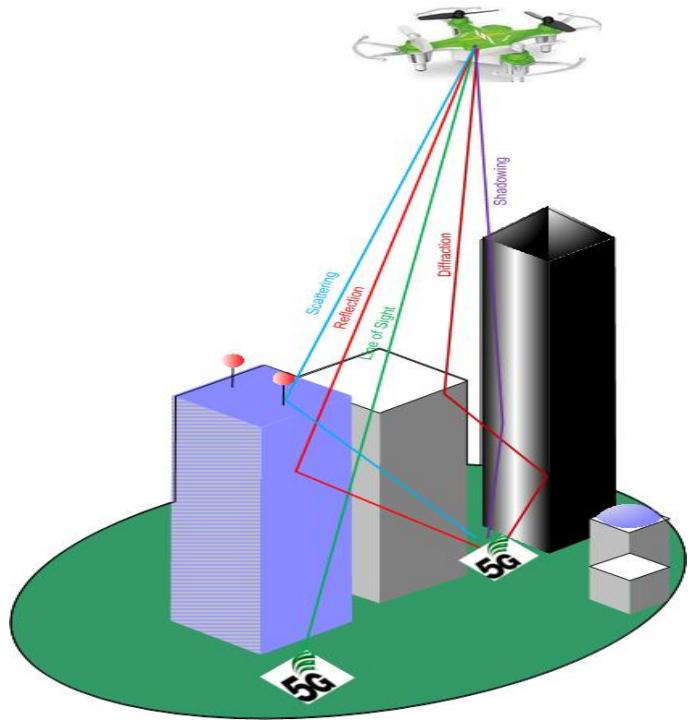

Fig.4 UAV Signal Propagation Mechanism

The signal strength depends on the distance, path loss and the height of drone. The signal path loss occurs because the signal took the multipath and received in different delays as shown in Fig.4. The signal path losses are varied according to the environmental impact. In an urban area, the path loss is very high because of a tall building. Moreover, it is small in suburban area and very small in rural area. The impacts of environment will destroy the signal by reflection, diffraction, and scattering, while as make signal weak. Reflection occurs when a propagating signal crash in large dimensions object. Diffraction occurs when a propagation signal path obstructed by the sharp surface in between the transmitter and receiver. Scattering happened when the propagation signal deviated from a straight path. Therefore, rough surface and small object lead to the scattered signal.

The communication channel between Drones and agents or IoT devices on the ground or ground center unit depends on some parameters such as fading, radiation power, and signal strength. Fading, signal strength and radiation powers depend on the coverage area and environment. The coverage probability and the sum rate for different IoT device deliver by UAV are discussed [35, 36]. Furthermore, deployment of UAV for optimal coverage of wireless communication introduced [37]. Several models have used for prediction the signal strength such empirical, analytical, theory, empirical and analytical and artificial intelligence models as shown in

Fig.5.

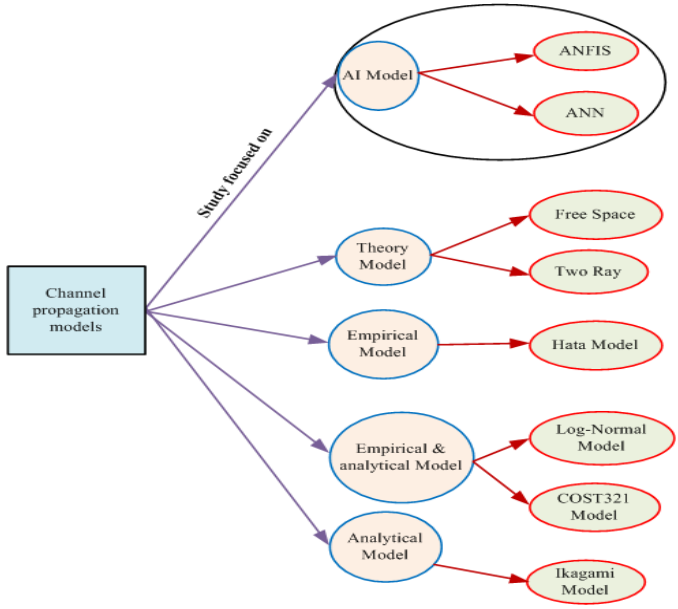

Fig.5 Channel propagation characteristics and Models

### 1. Empirical Rician channel fading prediction

The channel link is affected by multipath propagation, time dispersion, or fading. Urban area environment has a lot of big and tall buildings which cause the scattering, reflection, diffraction to the signal. Using drones will make communication channel link easier.

Fading occurred when attenuation suffered the modulation signal through the propagation media. Changing the position leads to time dispersion of the signal and time variance of the channel and this known by small and large scale fading. Several models are used for small-scale fading such as Rayleigh fading and Racian fading. Racian fading is taken into consideration for fading estimation in the case of LoS. The distribution of Rician is defined as:

$$f(r) = \frac{r}{\delta^2} e^{r^2+s^2/2\delta^2} I_0 \frac{rs}{\delta^2} \quad for\ r \gg 0 \qquad (1)$$

Where s is the field strength of the LoS component, and $I_0$ is the zero-order modified Bessel function of the first kind. K factor is the essential factor in the Rician channel which defined as:

$$K = \frac{s^2}{2\delta^2} \qquad (2)$$

When $K \to \infty$, the Rician PDF $\to$ Gaussian in the case of $K \to 0$, the Rician PDF $\to$ Rayleigh. Then, Rician PDF in regarding K-factor (dB) wrote:

$$f(r) = \frac{2r10^{k/10}}{s^2} e^{\frac{10^{k/10}}{s^2}(r^2+s^2)} I_0 \left(\frac{2r10^{k/10}}{s^2}\right) for\ r \geq 0 \qquad (3)$$

### 2. Empirical Signal Strength

Signal strength between the base station and mobile must be higher than the threshold value to maintain signal quality at the receiver [38]. Simultaneously, signal strength must not be powerful, because the strong signal will create co-channel interference with the channel in another cell which used the same frequency.

The received signal strength depends on path loss and transmitter and receiver parameters [39]. Quality of call establishment is based on received signal strength. Signal strength varies based on the environmental impacts and the average losses. On the one hand, necessary propagation models indicate that average RSS power decreases logarithmically with distance. On the other hand, path loss defines how much strength of the signal lost during propagation from the transmitter to receiver.

Propagation models predict the mean RSS for transmitter and receiver through separation distances as well as the variability of the signal strength in which a particular location is useful for predicting the radio coverage area of a transmitter and characterize signal strength over considerable separation distance between transmitter and receiver. Propagation models are helpful for predicting signal attenuation or path loss. The propagation path loss has a major limiting factor to coverage prediction which is derived from all losses encountered by the signal in its propagation from base station (BS) tower to the mobile user or mobile station (MS) [40]. The path loss information may be used as a controlling factor for system performance or coverage. The propagation path of HAP is investigated and analyzed in [23]. Hata model is available to predict the propagation loss. Hata model is an empirical path loss [dB].

$$PL1_{hata}(dB) = A + B\ log(d1) \qquad (4)$$

Where d is the distance in Km, A is fixed loss depends on frequency f in MHz and it is given by:

$$A = 69.55 + 26.16 log(f) - 13.82 log(h_b) - a(h_m) \qquad (5)$$
$$B = 44.9 - 6.55\ log(h_b) \qquad (6)$$

Where, $h_b$ is the height of base station antenna in m. $h_m$ is the height of mobile station antenna in meters. $a(h_m)$ is correlation factor in dBm and is given by:

$$a(h_m) = [1.1\ log(f) - 0.7]h_m - [1.56\ log(f) - 0.8] \qquad (7)$$

QoS defines the performance of any system and is used to assign a set of parameters which are proposed to signify assessable aspects of the subjective. Path loss plays a vital role in enhancing the QoS of wireless communication.

### 3. The probability of Line of Sight

Signal Line of sight means that the path between two fixed points is straight in the three-dimensional space. Environment impacts and elevation angle play a vital role to determine propagation features from UAV to IoT devices on the ground. For design a system, the transmitted signal should be received with high signal strength without distortions. The probability distortion for LoS corresponding to each elevation angle in each propagation environment was calculated in [18], the data has been created to be predicted. The LoS probability from smart UAV to IoT devices on the ground will be used to estimate the coverage area and capacity which allow IoT devices in the smart environments to move around and gather data under UAV coverage area.

The primary application of analytically of the channel is the ability to optimize the performance of the smart UAV rapidly. Altitude of the UAV plays a vital role in determines the coverage area and decrease the signal strength degradation on the ground. Increase the height of the UAV will lead to increase the probability of LoS which points to more objects/users in the LoS with UAV and higher path loss. Find the efficiently optimize; the altitude is significant for maximizing the probability of coverage and small losses. If the height of the UAV is h and the radius coverage area is R. Thus, the distance between UAV and the ground receiver is:

$$d = \sqrt{R^2 + h^2} \qquad (8)$$

Where the elevation angle of UAV concerning the user on the ground is:

$$\theta = tan^{-1}(\frac{h}{R}) \quad (9)$$

Then, the path loss is given by [17]:
$$PL(dB) = \begin{cases} 20\ log(4\pi f_c d/c) + \varepsilon_{los} & LoS \\ 20\ log(4\pi f_c d/c) + \varepsilon_{nlos} & NLoS \end{cases} \quad (10)$$

Where $\varepsilon_{los}$ and $\varepsilon_{nlos}$ are the average additional loss to the free space loss which depends on the environment, speed light, carrier frequency, and the distance between UAV and Iot devices on the ground.

Therefore, the LoS probability is crucial to predicting signal attenuation correctly among drone, IoT devices, robots or any objects on the ground. Prediction the accurate LoS probability determination helps to obtain more realistic path loss models. The elevation angle is the most critical parameter for predicting the LoS probability. Plotting the probability of LoS is depending on the parameters values of the environmental impacts $\alpha, \beta, \gamma$ respectively. The investigation of the probability of LoS condition based on [19] and the final formula of LoS probability is:

$$Plos = \prod_{n=0}^{m}[1 - e^{-\frac{[ht - \frac{(n+0.5)(ht-hr)}{m+1}]^2}{2\gamma^2}}] \quad (11)$$

$$m = r\sqrt{\alpha\beta} - 1 \quad (12)$$
$$r = h\ tan(\theta) \quad (13)$$

Where $hr, ht$ are the height of receiver and transmitter, respectively. r is the ground distance between receiver and transmitter. However, plotting the resulting base on (11) will smooth out because of the significant value of ht. Therefore, Plos is considered to be calculated based on $\theta$ and environment parameters impacts [18]. However, Holis et al. [18] directly suggested Plos as:

$$Plos = c1 - (C1-C2)/(1+(\frac{\theta-C3}{C4})^{C5}) \quad (14)$$

Where C1…..C5 are environments parameters which are given in [18]. However, Hourani et al. [19, 20] expressed the Plos by sigmoid term concerning elevation angle θ as:

$$P(LoS) = \frac{1}{1 + ae^{-b(\theta - a)}} \quad (15)$$

Where a and b are the S-curve parameters. Here, the Plos is easy to be calculated and analytical flexible.

4. *ANN*

ANN will process the data required for designing an appreciate UAV communication network. Therefore, we proposed to use ANN, which can process large date in very fast to find appreciate prediction and decision. ANN technique is aimed to maintain the connection among drone, IoT devices, robots on the ground and humans, effectively and efficiently. The ANN consists of some neurons arranged in a particular fashion as shown in fig.6. The input layer acts as an entry point for the input vector; no processing takes place in the input layer. The hidden layer consists of several Gaussian functions that constitute arbitrary basis functions (called radial basis functions); these basis functions expand the input pattern onto the hidden layer space. This transformation from the input space to the hidden layer space is nonlinear. The output layer linearly combines the hidden layer responses to produce an output pattern.

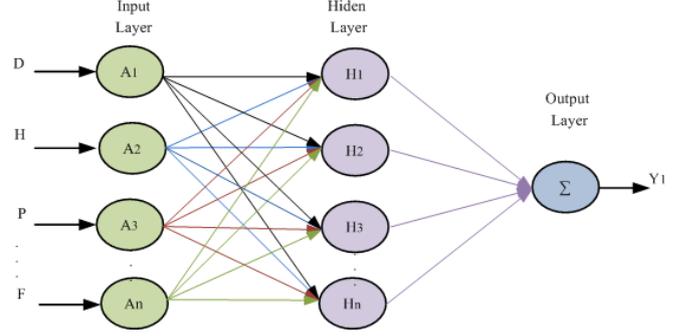

Fig.6 ANN training and testing the RSS

The inputs are including distance (D), altitude (H), frequency (F), and Path loss (P). The output equals the summation of the hidden layer. The output decides and estimates the appreciate signal strength (Y).

$$W_{k1}(n) = [W_k(n), … … … … , W_{k20}(n)] \quad (15)$$

Initialize al the following, the center value $\mu_{ji}(0)$, the span value $\delta_j(0)$, weight vector $W_K(0)$, expect $W_{11}(0) = W_{21}(0) = 1$. The calculation of the output of hidden layer, output and error are given respectively by:

$$Y_k = R[\sum_{I=I}^{M} W_{kj}(n)Z_J]\ , k = 1,2; M = 20 \quad (16)$$
$$e_k = d_k - y_k \quad (17)$$

Where $d_k \epsilon\{0,1\}$ desired pattern and update the weight given by:

$$W_{kj}(n+1) = W_{kj}(n) - \tau_w e_k z_j \quad (18)$$

Where $\tau_w$ and $\tau_\mu$ represent the learning rate of weight and center respectively, update the center and span momentum:

$$\mu_{ij}(n+1) = \mu_{ij}(n) + \tau_\mu \frac{z_j}{\delta_j}\left(x_i - \mu_{ji}(n)\right) \sum e_k w_{kj}(n) \quad (19)$$

$$\delta_j(n+1) = \delta_j(n) - \frac{2\tau_\delta z_j}{\delta_j(n)} ln z_j \sum e_k w_{kj}(n) \quad (20)$$

## V. Motivation and the proposed Technique

UAV is aerial robotics fly in the sky. It has attracted the interest for many applications in different environments significantly. UAV plays a vital role in wireless communication in some temporary events such as disaster, transport, marketing, monitoring emergency, traffic monitoring, sports, etc. Therefore, UAV also can perform intelligently different tasks in different fields for serve society, industry and government, better resolution image quality, low cost, rapid response, capable of flying without any condition, able to be closer to areas of interest dull, used in dirty and dangerous tasks. These are the motivation that attracted our attention to study the signal strength for keeping connectivity and coverage of UAV which can contribute to IoRT robots communication and its applications. To enhance the designing of the receiver and transmitter, signal strength should be predicted using an accurate technique. We proposed ANN for forecasting the precise signal and channel fading from different altitude and different distance. We proposed two scenarios for measuring the estimated signal strength from the UAV whether it fixed or in motion as shown in Fig.8 and Fig.8, respectively.

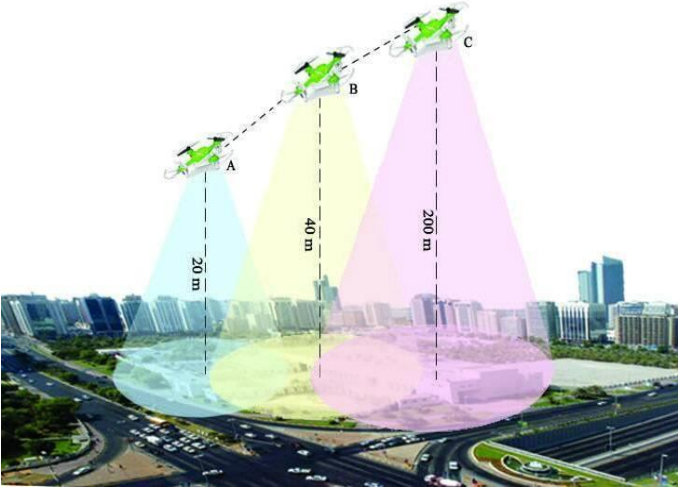

Fig.7 UAV signal strength from a different altitude

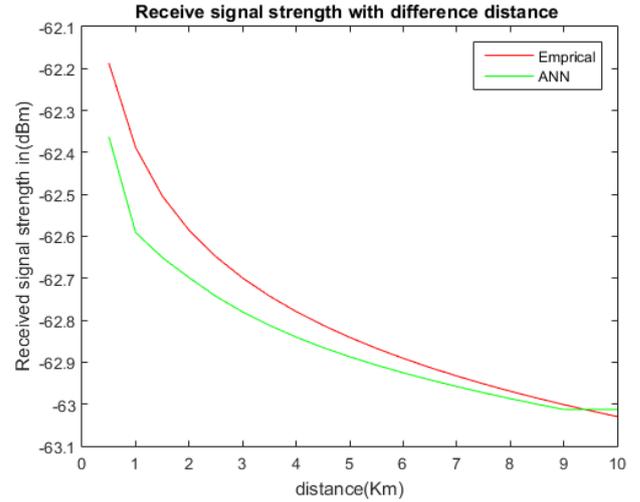

Fig.9 Signal strength prediction from different distances

In the first scenario, we consider the UAV is moving in different allocation in trajectory path to perform particular tasks. Each specific allocation of drone, we measured the signal strength. Fig. 6 shows the various allocation of UAV at A, B,…., H and I in different altitude 20 m, 40 m, …. , 180m and 200m respectively. Furthermore, the ANN is used to predict the signal strength and the channel fading in each particular allocation (altitude).

In the second scenario, the UAV altitude dynamically fixed and measures the signal strength in the various distances as shown in Fig. 7. ANN is applied for predicting the signal strength in a different position under the coverage of UAV technology.

The signal strength from the UAV with different UAV altitude is shown in Fig.10. The UAV was moving in different positions. Then, the desired QoS and signal connectivity in UAV coverage area estimated for each position. ANN applied for predicting the connectivity and coverage of UAV based on the signal strength with different altitude. Also, the signal strength predicted is appreciated by using ANN.

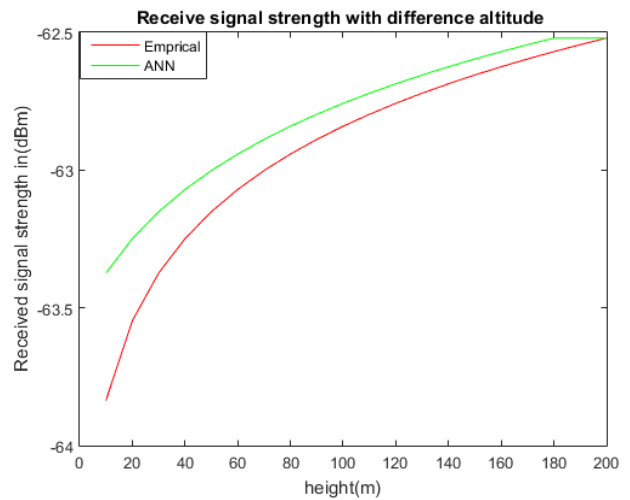

Fig.10 Signal strength prediction from a different altitude

The Rician channel fading has been selected for study the fading between the drones and other agents in the ground because Rician used for LoS. Fig.11 shows the Rician fading distribution with respected $K = 0, K = 50, K = 100$. The prediction of Rician fading is more accurate and appreciated.

The LoS probability Vs. Elevation angle was depicted in Fig.11 for urban and suburban, and dense urban area. The S curve in Fig.12 is agreed directly approximation to ITU-R P.1410-2, and the prediction was more accurate.

Fig.13 has depicted the LoS probability based on different elevation angle from the smart UAV for different environments. It is observed that the LoS probability in Fig.13 show similarities with LoS in [16] but here the prediction was more accurate because of the estimation of LoS probability at any location and in any angle.

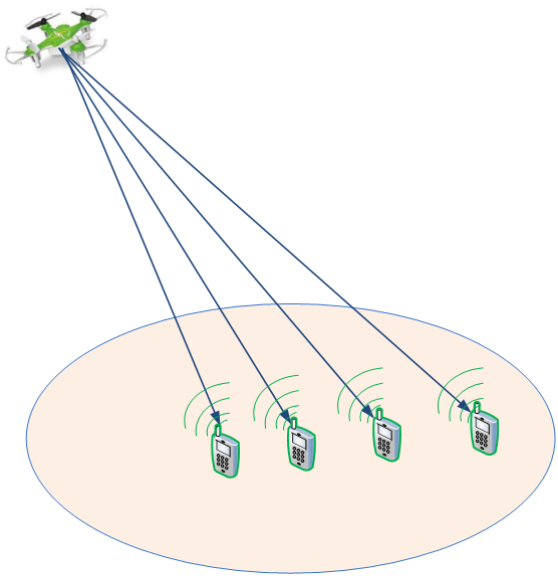

Fig.8 Signal strength in different distances

## VI. Result

Prediction is dynamic filter by using dynamic ANN for simulation monitoring, analysis, and control system varieties. ANN played a vital role in predicting appreciated received signal strength of desired QoS and connectivity for large coverage area which is closer to the agreement measurements.

The signal strength from the UAV is shown in Fig.9. The UAV was in a fixed position, and then the ANN applied for predicting the connectivity based on signal strength with different distance. It is shown that ANN predicts the signal strength significantly enhanced and better than the empirical method.

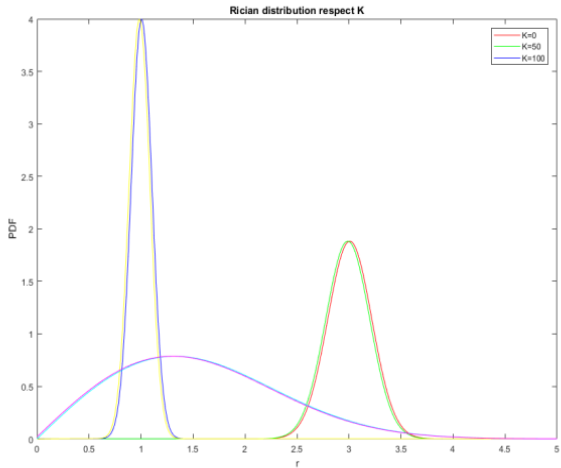

Fig.11 estimation of Rician channel fading

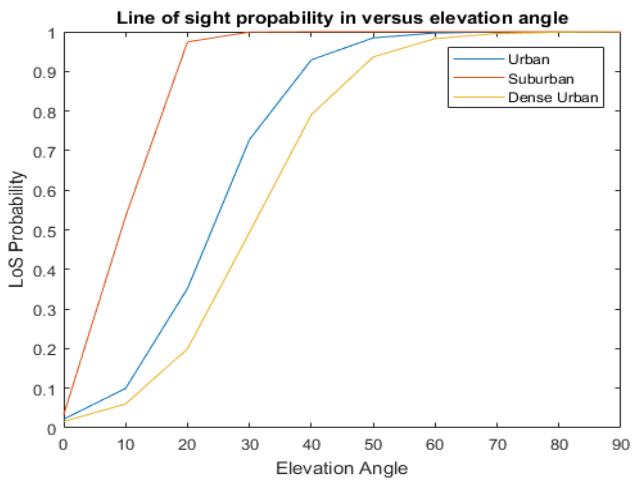

Fig.12 LoS probability of UAV for urban, suburban and dense urban

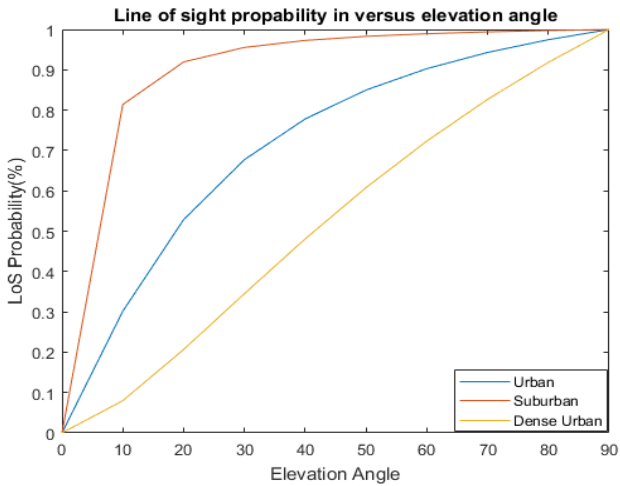

Fig.13 LoS probability depends on elevation angle in a different environment

## VII. Conclusion

Combination of artificial intelligent (AI) and the integration of UAV and IoT can produce amazing sophisticated solutions to today's problems. Therefore, it is considered a microcosm of IoT and already beginning to replace that connected sensors at stationary efficiently. IoT devices are unable to send the capture data to long distance. Furthermore, integration of UAV and IoT devices will bring a solution to our today problems. Smart UAVis proposed here for gathering data for different IoT device and makes a decision and then sends data to the final destination. Therefore, smart UAVconsider as UAV as services. The signal strength from smart UAV is affected by environmental impacts. Therefore, it is difficult to achieve high accuracy for estimating the signal strength, if unpredictability of channel makes between UAV or IoT devices. As long as, it is required for design the appreciate transmitter and receiver according to the ecological effects. Then, the signal will be easy to recover at the receiver. Design the receiver depends on the accurate prediction of the signal. The efficient and accurate signal strength is predicted based on altitude, path loss, channel distribution, elevation angle and the distances, using ANN is analyzed. ANN plays an essential role in training big data in very fast. Furthermore, it gives a correct prediction and appreciates decision. Notably, it is relatively difficult to find a method of signal strength estimation which can achieve a general estimation. Because of the signal strength depends on the wireless channel dynamic variation terrain characteristics. However, the ANN can give an accurate prediction where a very close agreement to generic estimation.